\documentclass{elsart} 
\usepackage[english]{babel} 
\usepackage{times}
\usepackage[iso]{umlaute} 
\usepackage{epsfig} 
\usepackage{setspace}
\usepackage{amsmath}
\begin{document} 
\begin{frontmatter}
\title{Modeling of Damage Evolution in Soft-Wood Perpendicular to Grain by means of a Discrete Element Approach} 
\author[wittel]{Falk K. Wittel\corauthref{cor}},
\corauth[cor]{Corresponding author. Tel.: +49-711-685-7093; fax: +49-711-685-3706}
 \ead{wittel@isd.uni-stuttgart.de}
\author[fmpa]{Gerhard Dill-Langer},
\author[wittel]{Bernd-H. Kröplin}

\address[wittel]{Institute for Statics and Dynamics of Aerospace Structures, University of Stuttgart, Pfaffenwaldring 27, 70569 Stuttgart, Germany}
\address[fmpa]{Otto-Graf-Institute (MPA), University of Stuttgart, Pfaffenwaldring 4, 70569 Stuttgart, Germany}

\begin{abstract}
The anisotropy of wood within the radial-tangential (RT) growth plane has a major influence on the cracking behavior perpendicular to grain. Within the scope of this work, a two-dimensional discrete element model is developed, consisting of beam elements for the representation of the micro structure of wood. Molecular dynamics simulation is used to follow the time evolution of the model system during the damage evolution in the RT plane under various loading conditions. It is shown that the results are in good agreement with experiments on spruce wood, and that the presented discrete element approach is applicable for detailed studies of the dependence of the micro structure on mesoscopic damage mechanism and dynamics of crack propagation in micro structured and cellular materials like wood.
\end{abstract}
\end{frontmatter}
\section{Introduction} 
Wood is probably the most ancient structural material in the world, widely used today in many species, for all kinds of purposes, leading to a world production of roughly $10^9$ tonnes, which is about even to the one of iron and steel \cite{gibson-ashby-97}. It's application in modern civil constructions of large dimensions like stadium roofs or long-span bridges calls for a good understanding of design properties like moduli, crushing strength and toughness.  Wood is a natural fiber composite, which is strongly orthotropic due to its internal material structure. The elemental material is a cellulose cell wall material with equal properties for different species \cite{gibson-ashby-97}. Nevertheless, macroscopic properties vary highly within the tree, among different species and along loading directions. Explanations for these extreme differences are found on the microstructural level in the cellular structure. Several internal length scales can be found, that are relevant for the damage evolution, including many defects. Depending on the observation scale and position, wood shows different physical and geometrical properties. The size scales are typically subdivided into hierarchical levels like the atomic, micro, meso and macro scale. The material can be regarded as a cellular structure with hexagonal shaped wood fibers arranged in parallel on the micro scale (comp. Fig.1). The dominating structure on meso scale in the RT plane, perpendicular to the tree axis, are annual rings with an alternating early- and latewood that differ in fiber shape and thickness of the cell walls. For the understanding of damage in wood, damage mechanisms on the micro and meso scale are of special interest.

\begin{figure}[htb] 
  \begin{center}
    \epsfig{bbllx=19,bblly=19,bburx=576,bbury=738,file=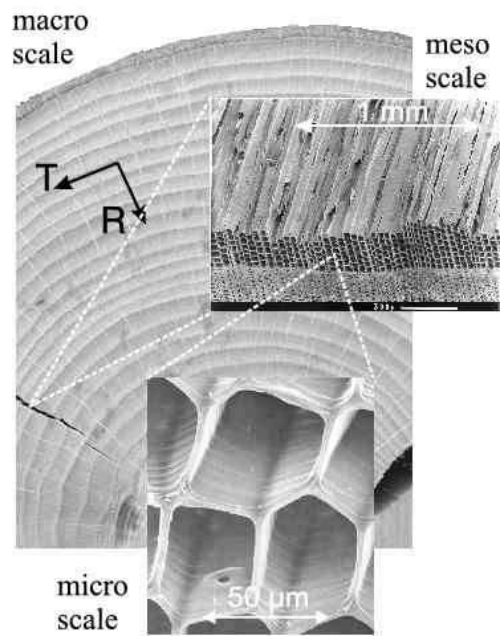, width=6.5cm} 
    \caption{Relevant scales in wood (Micro scale picture from \cite{meylan-butterfild-72})}
  \end{center}
\end{figure} 

Discrete Element Methods (DEM) are a fast emerging computational method designed to solve problems with gross discontinuous materials and geometrical behavior. Due to the creation and continuous motion of evolving crack surfaces, fracture of material is difficult to handle numerically. Continuum models have problems to account for the discrete nature of material failure. One approach is a smeared crack approach, e.g. with a micro-plane damage model \cite{ozbolt-aicher-2000}, where the influence of the damage is considered in terms of the material properties. The other approach is to continuously change the model topology, as cracks evolve \cite{lucena-simon-etal-2000}. Alternatively, discrete models like lattice dynamics models can be used to simulate fracture \cite{herrmann-roux-90}. A common type of question around DEM is related to the possibility of assigning continuum properties to a given lattice structure, especially avoiding artifacts e.q. due to a regular micro structure of a model. Contrary, for the representation of wood, a regular microstructure like a hexagonal lattice is especially suited for naturally taking the anisotropy of wood in the RT plane into account. 

In the present paper fracture of wood was studied with a combined beam-particle lattice model on the micro- and mesoscopic scale. The mesoscopic structure of annual rings is projected on the microscopic model and generic failure mechanisms, observed in the failure of spruce wood in the RT plane, are implemented into the model. The paper is organized as follows: Section 2 gives a detailed description on experimental in-situ studies of the fracture processes in clear softwood under various loading conditions perpendicular to grain. Experiments are focused on the micro-meso-structural scale, using a confocal laser scanning microscope (CLSM). Identified generic damage mechanisms are the rupture of earlywood cell walls for crack propagation in tangential direction and debonding of the interface of wood fibers for crack propagation in radial direction. Section 3 provides an outline of the theoretical background of the discrete element model, for the micro-structure of wood. After the model description, test simulations and damage simulations are presented and compared to the experimental findings in Section 4.
\section{Damage in wood in the RT-plane}
Damage basically initiates on the atomic scale and reaches relevance for larger scales like the micro or meso scale while it propagates, leading to global failure when reaching the macro scale. For the prediction of the damage evolution in materials scale dependent knowledge on their relevance for the damage evolution is indispensable and intimately connected to its characteristic structure. The typical observation scales of wood are shown in Fig.1. Looking at the sample of wood sufficiently distant from the center of the tree, three orthogonal planes of symmetry are found: radial $(R)$, tangential $(T)$ and longitudinal $(L)$. Stiffness and strength are greatest in longitudinal direction. In real constructions load perpendicular to grain is often present and damage consequently. Macroscopic material values for spruce wood are given in the table \cite{lucena-simon-etal-2000,wood-handbook}. Even though wood, as we use it, is just the dead corpus of a living tree, it is the result of its process of formation. Therefore, deviations within one board of wood can be in the range of $25\%$ \cite{wood-handbook}, since during the growth process, trees have to react to environmental conditions by structural adaptation on all scales. 

\begin{tabular}{l} 
Typical properties for spruce wood \cite{lucena-simon-etal-2000}:\\
{\sc Young}'s / shear moduli $[MPa]$ / {\sc Poisson}'s ratios:\\
\begin{tabular}{p{1cm}|p{1cm}|p{1cm}||p{1cm}|p{1cm}|p{1cm}|p{1cm}|p{1cm}}\hline
$E_R$ & $E_T$ & $E_L$ & $G_{RT}$ & $G_{LR}\approx G_{LT}$ & $\nu_{RT}$ & $\nu_{LR}$ & $\nu_{LT}$\\
$1200$ & $800$ & $15000$ & $50$ & $700$ & $0.38$ & $0.25$ & $0.38$\\ \hline 
\end{tabular} 
\end{tabular}

On the meso scale, wood is a cellular solid with mainly cells of hexagonal cross-section (also called tracheids in the case of soft-wood). The characteristic annual rings in soft-wood consist of alternating circumferential bands of thick- and thin- walled tracheids. Fig.2 shows new measurements of the relative density distribution $\rho^*/\rho_c$, based on CLSM pictures for an annual ring with the cell wall density $\rho_c$. Other features of the meso scale are rays and sap channels, that are important for the water balance and growth of trees but are of minor importance for the material representation of soft-wood \cite{gibson-ashby-97,lucena-simon-etal-2000,persson-97}. Of course, structural differences between species are significant. Micro mechanical finite element models reveal the cell density to be the governing parameter of the macroscopic mechanical properties of wood within the RT plane \cite{persson-97}.
\begin{figure}[htb] 
  \begin{center}
    \epsfig{bbllx=19,bblly=19,bburx=576,bbury=502,file=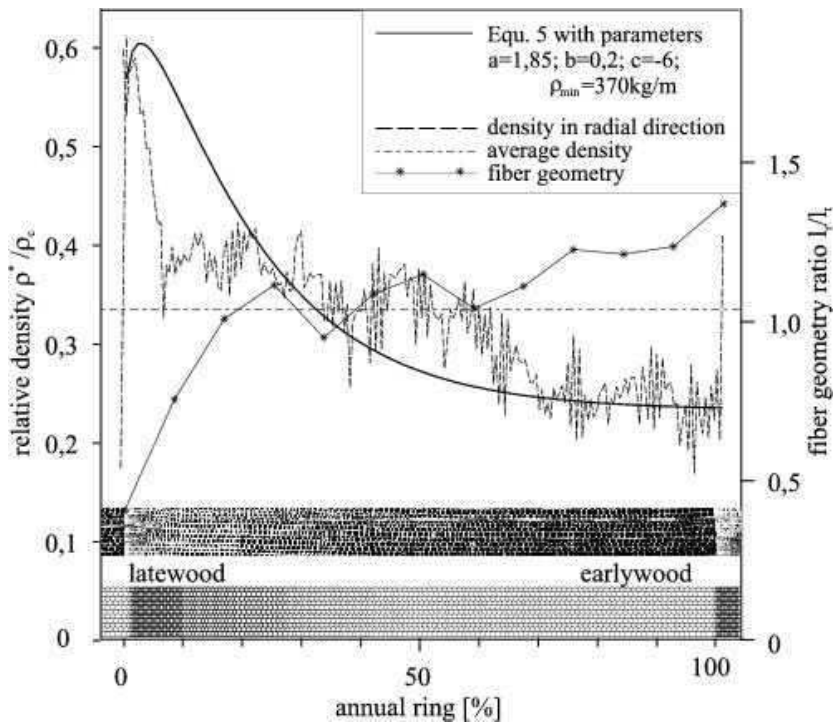, width=11cm} 
    \caption{Optically measured relative density, cell aspect ratio and microstructure in one annual ring} 
  \end{center}
\end{figure}

Fibers are glued together by lignin, leading to a natural fiber composite of less than 5\% matrix volume fraction. On the finer micro scale, wood is again a fiber-reinforced composite, since the cell walls are made up of fibres of cellulose, which is a crystalline polymer with sort amorphous regions, embedded in an amorphous hemi-cellulose and lignin matrix \cite{persson-97}. These micro-fibrils are oriented in layers of different orientation, depending on the time of formation \cite{gibson-ashby-97,wagenfuehr-88,meylan-butterfild-72}. 
\begin{figure}[htb] 
  \begin{center}
    \epsfig{bbllx=19,bblly=19,bburx=576,bbury=473,file=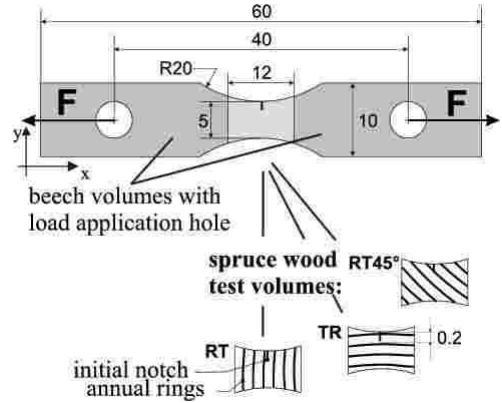, width=6.5cm} 
    \caption{Specimen with dimensions in $mm$ and different investigated configurations of load to growth directions} 
 \end{center}
\end{figure} 

\begin{tabular}{l} 
Cell wall properties for softwood \cite{gibson-ashby-97}:\\
{\scriptsize
\begin{tabular}{llllllll}
cell wall density& $\rho_c$ &$[kg/m^3]$&  $1500$  & Tracheid size longitudinal&$l_L$&$[mm]$&$2,5-7,0$\\
{\sc Young}'s modulus(T) &$E_T$& $[GPa]$&$10$& tangential size &$l_T$&$[\mu m]$&$30$\\
Shear modulus & $G_T$& $[GPa]$ & $2,6$ & radial size &$l_R$&$[\mu m]$&$12-40$\\
Yield strength (T) & $\sigma_c$& $[MPa]$ & $50$ & wall thickness &$t$&$[\mu m]$&$1,6-5$\\
Volume fraction & &$[\%]$ &$85-95$ &&&&\\
\end{tabular}}
\end{tabular}

Even though different kinds of soft wood have a large difference in relative density, the cell wall properties for all species are pretty much identical. This is reasonable, since in nature, optimization of structural materials is only possible in the framework of feasible biological processes, which are identical for all trees. Therefore optimization of microstructure and shape is indispensable for survival. To understand the mechanical and breakdown behavior of wood, knowledge on the material-structure composition of the high strength elements of wood - the cells is necessary.
\subsection{Microscopic in-situ experiments}
The damage evolution in wood on the micro-meso structural level in the RT-plane was studied under tension. All test specimen are clear spruce wood, cut from one board with a mean density of $500kg/m^3$ and a moisture content of $12\%$. The annual ring width $d_{jr}$ varied between $1mm$ and $2.5mm$ for different specimen, due to different locations within the board cross section. Details on specimen preparation and experimental setup are published in \cite{dill-langer-aicher-2000,dill-langer-luetze-aicher-02}. 

\begin{figure}[htb] 
  \begin{center}
   \epsfig{bbllx=19,bblly=19,bburx=576,bbury=329,file=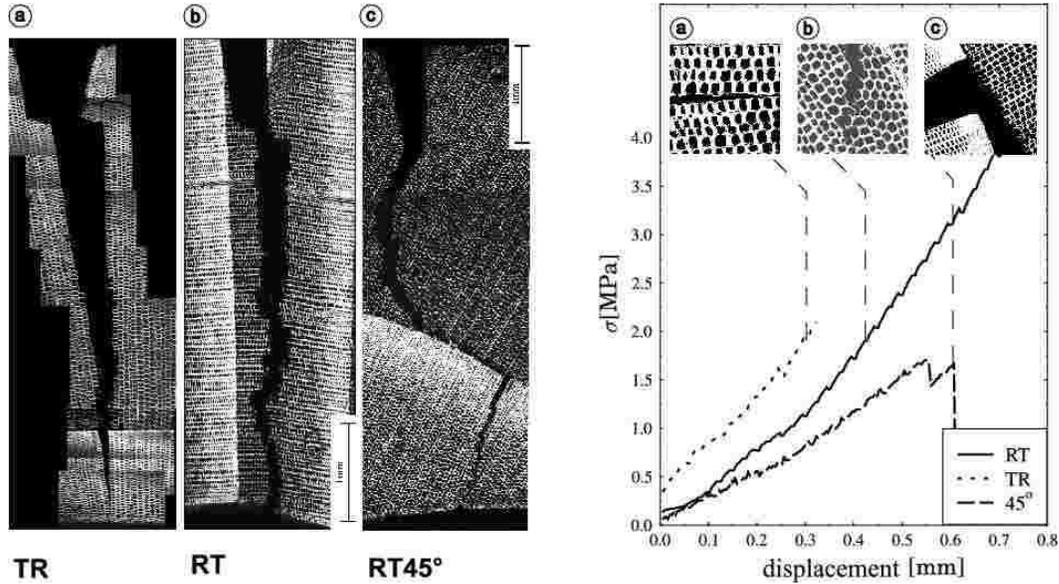, width=14.0cm}
 \caption{Comparison of crack patterns for different configurations $(top)$ and load displacement curves for $TR,RT$ and $RT45^o$-configuration $(bottom)$.} 
 \end{center} 
\end{figure}

In-situ tests are made using a Confocal Laser-Scanning Microscope (CLSM), therefore intensities on the pictures represent height values from a focus plane.
\subsection{Micro- and mesoscopic degradation of wood}
Different orientations lead to completely different damage evolutions. If wood is loaded in tangential direction ($TR$-specimen (Fig.4$a$)) the crack propagates straight in radial direction with minor deviations and smooth crack surfaces. For the $RT$ configuration (comp. Fig.4$b$) the crack propagates within the earlywood parallel to the annual ring and perpendicular to the loading direction. The crack surface is rough with edges and steps and the crack propagates mostly in an instable manner. Compared to pure $RT$ an $TR$ crack configurations a zig-zag behavior is observed for orientations under $45^o$ ($RT45^o$ Fig.4$c$). Initiated at the notch, the crack first propagates in radial direction with smooth crack surfaces. While the crack approaches the early/latewood transition, it is deflected from the initial direction until it propagates along the transition edge. The final failure occurs, when half of the specimen is damaged, by a sudden crack formation in radial direction. 

Microscopic observations reveal different damage mechanisms. Tension stress in radial direction is mainly released by rupturing the cell walls, leading to rough and stepped surfaces (comp. Fig.4$bottom$ insert $b$) while for tension stress in tangential direction cells are generally intact, but inter-cellular damage (fiber debonding) is the dominant damage mechanism (comp. Fig.4$bottom$ insert $a$). For the $RT45^o$ configuration we find a combination of inter- and intracellular damage (comp. Fig.4$bottom$ insert $c$), depending on the direction of the crack propagation.
Further details on the damage evolution under tension can be found in \cite{dill-langer-aicher-2000,dill-langer-luetze-aicher-02}.
\section{A discrete element model for wood} 
In order to study the damage evolution of wood, a two-dimensional network model is worked out. Molecular Dynamics (MD) simulation is used to follow the motion of each cell by solving Newton's equations of motion. In the present study we use a $4^{th}$ order {\sc Gear} Predictor Corrector scheme \cite{allen-tildesley-87}. A general overview of MD simulations applied to composite materials can be found in \cite{herrmann-roux-90}.  
 \begin{figure}[htb] 
  \begin{center}
    \epsfig{bbllx=19,bblly=19,bburx=576,bbury=279,file=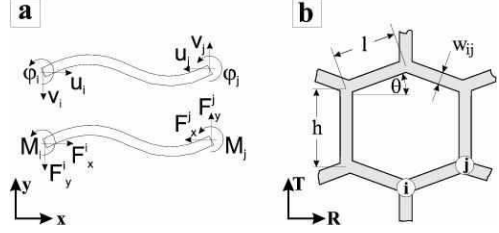, width=6.5cm} 
    \caption{Degrees of freedom and internal loads on a single beam element and hexagonal lattices.}  
  \end{center}
\end{figure}

With this method already small cracks are sharply defined, with the possibility to simulate simultaneously a conglomerate of cracks within rather large lattice sizes \cite{herrmann-roux-90}. The fundamental advantage of the lattice model used in this investigation is due to its simplicity, giving direct access and possible physical interpretation to each step of the algorithm. Consequently, one can modify the rules for features of interest, like the characteristic properties of size, strength or force in a rather straightforward and transparent way.
 
The model is composed in four major steps, namely, {\em a)} the implementation of the microscopic structure, {\em b)} the determination of the constitutive behavior, {\em c)} the breaking of the model and finally {\em d)} the construction of the test specimen.
 
{\bf{\sl (a) Micro structure: }} 
The basic structural element used is a two-dimensional {\sc Timoshenko} combined spring-beam element for small displacements, in other words elements, that can elongate like springs and bend like beams at the same time. An overview over network models is given in \cite{ostoja-starzewski-02}. The nodes with mass $m_i, i=1...N$ of a two-dimensional lattice with $N$ nodes are connected via these elastic elements. An element connecting the positions $i$ and $j$, has the cross section $A^{ij}$ and the geometrical moment of inertia $I^{ij}$ of the beam for flexion. The length $l^{ij}$ of all beams is equal for this case. The elastic behavior is also governed by the two material dependent constants for the {\sc Young}'s- and shear moduli of the beam $E_b$ and $G_b$. For a beam between sites $i$ and $j$ the normal, shear and bending flexibilities $a_{ij}, b_{ij}, c_{ij}$ are given by:
\begin{equation}
a_{ij}=\frac{l_{ij}}{E_b A_{ij}}, \qquad b_{ij}=\frac{l_{ij}}{G_bA_{ij}}, \qquad  c_{ij}=\frac{l_{ij}^3}{E_bI_{ij}}.
\end{equation}

Each node $i$ has in the global coordinate system the position $x_i, y_i$ and the rotation $\theta_i$ and all the forces and moments of the beams, connected to $i$ act on it, allowing only motion in the observation plane (comp. Fig.5$a$). In the local coordinate system of the beam, three continuous degrees of freedom are assigned to both ending points $i,j$ of the beam, which are for site $i$ the displacements $u_i$ and $v_i$ and the bending angle $\phi_i$. The longitudinal force acting at site $i$ is
\begin{equation}
F_{b,x}^i = \alpha_{ij}(u_j-u_i),   \quad \text{with} \quad \alpha_{ij}=1/a_{ij},
\end{equation}
the shear force is
\begin{equation}
F_{b,y}^i = \beta_{ij}(v_j-v_i)-\frac{\beta_{ij}l_{ij}}{2}(\phi_i+\phi_j), \quad\text{with} \quad  \beta_{ij}=1/(b_{ij}+c_{ij}/12),
\end{equation}
and the flexural torque at site $i$ is
\begin{equation}
        M_{b,z}^i=\frac{\beta_{ij}l_{ij}}{2}(v_j-v_i+l_{ij}\phi_j)+ \delta_{ij}l_{ij}^2(\phi_j-\phi_i),  \quad \text{with}  \quad \delta_{ij}=\beta_{ij}(b_{ij}/c_{ij}+1/3),
\end{equation}
consequently resembling the {\sc Timoshenko} beam theory. Quantity $b_{ij}$ was chosen to be $b^{ij}=2a^{ij}$, resulting in a {\sc Poisson}'s ratio of $\nu_b=0$. The model is derived from a model for geomaterials, presented in \cite{kun-herrmann-96,kun-herrmann-2001}. For simplification purposes, the beams behave in a time independent way, but time dependent beam properties are easy to introduce. In our simulation, the nodes are the smallest particles interacting elastically with each other.

\begin{figure}[htb] 
  \begin{center}
    \epsfig{bbllx=19,bblly=19,bburx=576,bbury=693,file=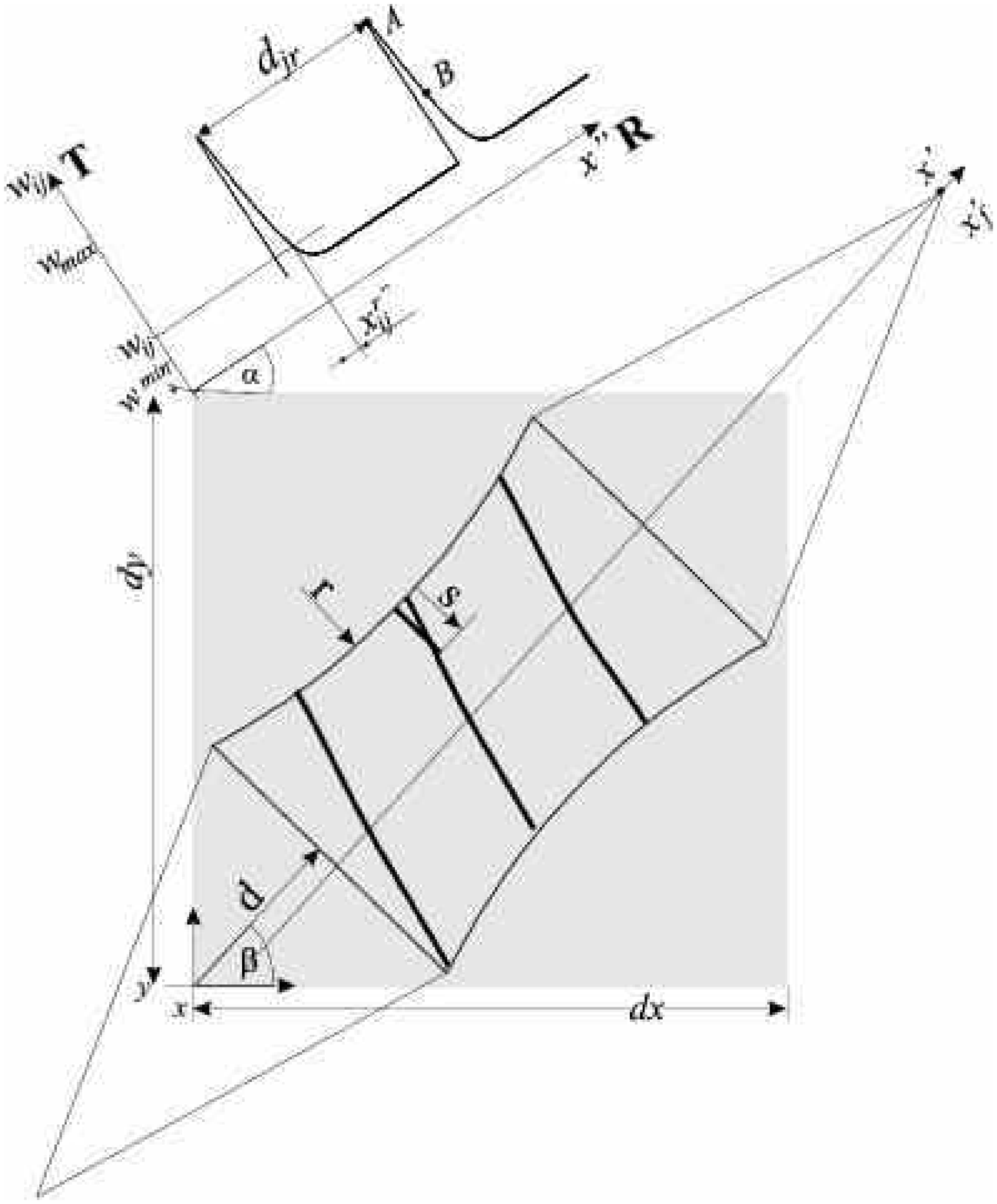, width=8.5cm}
    \caption{Construction of the model with ground lattice, meso structure projection and cutting of the specimen}
  \end{center}
\end{figure}
The construction of the final model is done in several steps shown in Fig.7. First a hexagonal beam lattice is constructed. As second step, the meso structure of the annual rings is projected on the fundamental lattice with the possibility to vary the distance $d_{jr}$ and angle $\alpha$ of the annual rings with respect to the specimen axis, which is identical to the loading direction. The different material properties of an annual ring are represented by alternating the effective density $\rho(x^{ij''})=\rho^*$ according to the function
\begin{equation}
        \rho^*=\rho_{min}\left(1+2 \cdot a x_r^{ij''^b}\cdot e^{cx^{ij''}_r}\right)
\end{equation}
representing an annual ring. If the sample of wood is cut distant from the center of the tree, the curvature of annual rings can be neglected, leading to orthotropic material properties. Eq.5 is used in order to obtain a steady density function. Dimension $x_r^{ij^{''}}$ stands for the distance of the middle point of the beam from the annual ring interface, connecting sites $i$ and $j$ in relative coordinates of the $R-T-$material coordinate system (compare Fig.7). The parameters $a,b,c,$ are selected in a way, that the properties scale according to the density between an annual ring (comp. Fig.2). The thickness for the beam connecting site $i$ with $j$ depends on the lattice type and parameters chosen. 

The required relationship between the cell wall thickness $w_{ij}$, the edge lengths $h$ and $l$, and the relative density $\rho^*/\rho_c$ for hexagonal lattices with corrections for overlapping areas is
\begin{equation}
w_{ij} ( \rho^* )= A \cdot \left( (h+2l) -  \sqrt{(h+2l)^2 - \cos \theta (2h+l \sin \theta) \frac{2l\rho^*}{A\rho_c} }\right)
\end{equation}
with $A=1+\sin \theta + \cos \theta$.

{\bf{\sl (b) Constitutive behavior: } }
Since the model is of two-dimensional nature, we analyze its response to load applied in the $R-T-$plane. Even though all elements are considered to be isotropic, anisotropy arises due to their arrangement and the macroscopic moduli $E^*_R,$ $E^*_T$, $G^*_{RT}$, $\nu^*_{RT}$ (since $E^*_R \nu^*_{TR}=E^*_T \nu^*_{TR}$ is assumed) and plateau values for the compressive stresses $\sigma^*_R$ and $\sigma^*_T$ if compressive failure is dominant, are required to describe the mechanical behavior of the composite. The derivation for the relations assuming linear elastic deformations can be found in \cite{gibson-ashby-97}.

For regular hexagonal lattices with the parameters $h$, $l$, $\theta$ and the beam thickness $w$ the {\sc Young}'s moduli $E^*_T$ and $E^*_R$ and the shear modulus $G^*_{RT}$ are calculated with the properties of the cell wall $E_c$ as
\begin{equation}
\begin{split}
\frac{E^*_R}{E_c}=\left( \frac{w}{l}\right)^3 \frac{\cos \theta}{(h/l+\sin \theta)\sin^2 \theta};\qquad \frac{E^*_T}{E_c}=\left( \frac{w}{l}\right)^3 \frac{(h/l + \sin \theta)}{\cos^3 \theta};\\ \frac{G^*_{RT}}{E_c}=\left( \frac{w}{l}\right)^3 \frac{(h/l + \sin \theta)}{(h/l)^2(1+2h/l)\cos \theta}; \qquad \nu^*_{RT}=\frac{\cos^2 \theta}{(h/l+\sin \theta)\sin \theta}.
\end{split}
\end{equation}
With these expressions it is possible to estimate the correct values for the microstructure iteratively, since a change of any property also effects the other properties through the density function Eq.5. Good microstructural values for spruce wood were found to be $E_c=11GPa$, $h=1.5l$ and $\theta=12°$, but of course these values depend on the number of wood fibers, that form a annual ring, leading to cell wall thicknesses $2.6\mu m \le w_{ij} \le 10.3 \mu m$. These values do not fit the observed microstructure, since Eqs.7 are only valid for regular hexagonal lattices. Solutions for the real model are presented in Section 4.

{\bf{\sl (c) Breaking of the material: } }
Stresses inside the system can be released by the fracture of the beam elements, either under tension or under bending. In the framework of Discrete Element Methods, the complicated crack-crack interaction is naturally taken into account. If the deformation of an element due to the total forces acting on it exceeds its breaking thresholds (damage thresholds comp. Fig.6), its stiffness is abruptly reduced to zero, resulting in a load redistribution to neighboring beams in the next iteration steps. If after some iterations due to the load redistribution to the neighboring beams they exceed their threshold value due to the additional load, they fail too, giving rise to crack growth. Alternatively, so called continuous damage laws with incremental stiffness reduction, once the actual damage threshold is exceeded, can be used \cite{raul-feri-hans-01}. Since we have no particular basis for a topological disorder or cell property disorder as well, we make a computational simplification by assuming the disorder in beam properties to adequately account for all the relevant disorder present in the represented material region. Disorder is introduced to the model by a two parameter Weibull distribution of breaking thresholds $\varepsilon_d$ at the beginning of the simulation and is kept constant during the fracture process (quenched disorder) as
\begin{equation} 
P(\varepsilon_0)=1-\exp\left(\left(-\frac{\varepsilon_d}{\varepsilon_0}\right)^m\right),
\end{equation} 
with the Weibull modulus or shape parameter $m=6$ and the scale parameter $\varepsilon_0=0.0055$.
\begin{figure}[htb] 
  \begin{center}
    \epsfig{bbllx=19,bblly=19,bburx=576,bbury=323,file=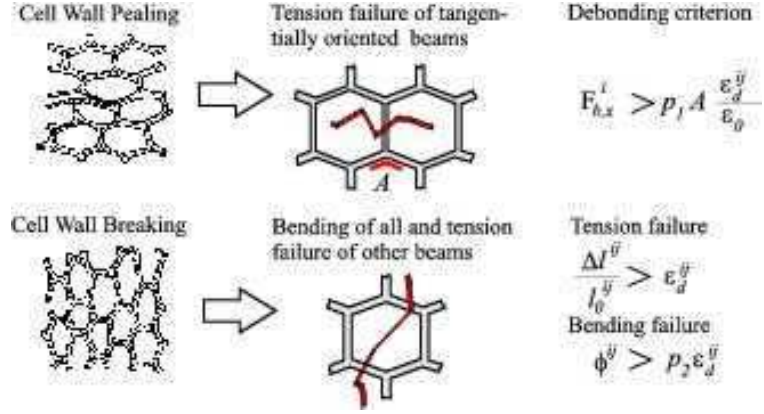, width=10cm}
    \caption{Representation of different damage mechanisms} 
  \end{center}
\end{figure} 

The two experimentally identified stress release mechanisms, cell wall breaking and peeling are represented in different ways. The peeling mechanism (comp. Fig.6) is activated, when the force on a node in tangential direction exceeds a force threshold value. This value is calculated by multiplying a critical debonding stress $p_1$ and the area $A$ affected by the debonding, since the matrix material is the same for the whole wood material (comp. Fig.6). Contrary, the cell wall breaking depends on the cell wall thickness. The breaking criteria always use the same disorder for all damage mechanisms for simplicity. $p_2$ is a parameter to take the quantitative relation of the critical values for tension and bending into account.

{\bf{\sl (d) Construction of the specimen: } }
Finally, the simulation model is extracted out of the basic hexagonal lattice under the angle $\beta$ by deleting those elements, which do not fit in the specimen geometry given by Fig.3. As a last step, the border elements are identified and connected to the two loading points via elastic beams. This way, the specimen is free to rotate as observed in the experiments, when the crack propagates.
\section{Computer simulations}
After the model construction, computer simulations for the constitutive behavior of the material are performed. This is done by calculating the time evolution of the particle assembly by solving the equation of motion of the nodes in terms of molecular dynamics, 
\begin{equation} 
  m_i\ddot{\vec{r}_i} = \vec{F}_i, \qquad i=1, \ldots , N 
\end{equation} 
where $N$ denotes the number of nodes, $m_i$ is the mass and $\vec{F}_i$ is the total force acting on node $i$. A $4^{th}$ order Predictor Corrector algorithm is used in the simulations to solve numerically the second order differential equation system (Eq.9) \cite{allen-tildesley-87}. After each integration step the breaking condition is evaluated for all the intact beams. Those beams which fulfill this condition are removed from the further calculations. The discrete element method gives the possibility to monitor the development of the microscopic damage in the specimen.
\begin{figure}[htb]
  \begin{center}
    \epsfig{bbllx=19,bblly=19,bburx=470,bbury=470,file=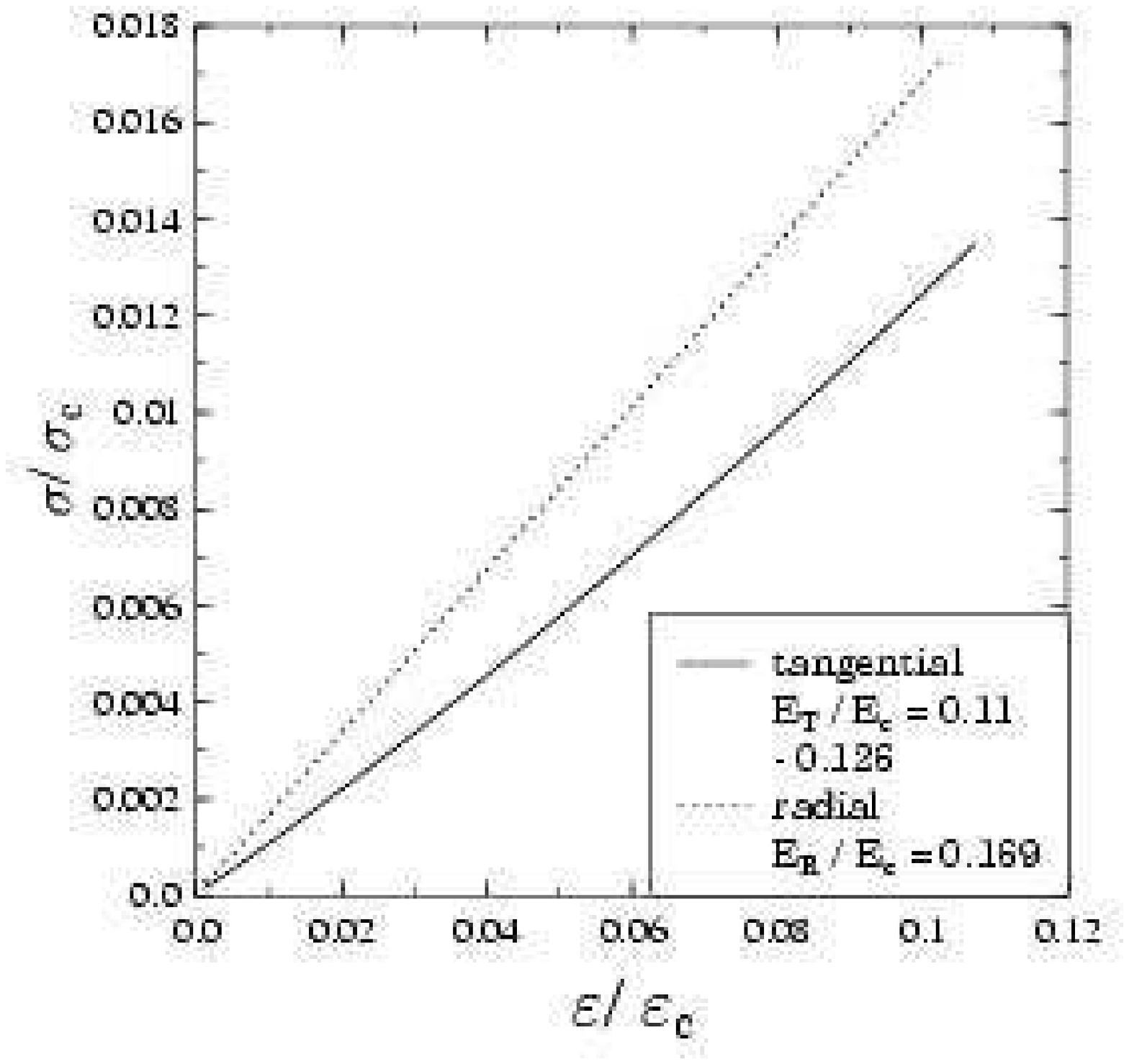, width=6.5cm}
    \caption{Stress-strain curve for a sample of $1x1mm$ with annual ring distance of $1mm$}
  \end{center}
\end{figure} 

Material properties are measured numerically on the macroscopic scale by computer simulations on the bulk model. All beams are considered to be linear elastic up to failure, so no plastic or time dependent creep behaviour is considered in this stage of the analysis, but basically the elements can easily be replaced by other kinds of rheological elements. Fig.8 shows the constitutive behavior of a model block with different material orientations. Values are transformed into non-dimensional form for comparison, using characteristic values of single beams . The increase of the radial {\sc Young}'s modulus $E_R$ is due to the alignment of the beams in loading direction.

The {\sc Young}'s modulus is calculated via the energy $e_{pot}$ stored in the beams as $E_{R/T}=2e_{pot}/V \cdot \varepsilon^{-2}$ for the volume $V$, and is in accordance with values for spruce wood.

As a next step, breaking is switched on and the three configurations $RT/TR/45^oRT$ are calculated. Activated crack mechanisms and crack evolution are in agreement with the experimental observations.
\begin{figure}[htb]
  \begin{center}
    \epsfig{bbllx=19,bblly=19,bburx=576,bbury=174,file=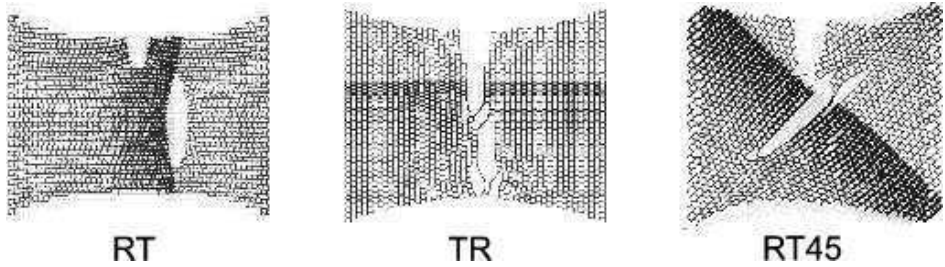, width=12.5cm}
    \caption{Microstructure of damage for the three different configurations $RT;TR;45^oRT$} 
  \end{center}
\end{figure}  

As a further application, we simulate the permeation of a circular object (e.g. a nail) in the center of our model by defining a repulsive contact force. During the simulation we constantly increase the diameter of the nail, leading to damage. Contact forces are only calculated for the interaction of the nail with the cell structure and not within the cell structure itself. We observe crack growth mainly in the tangential direction. A detailed study on the compressive failure of cellular materials with the discrete element method will be published in a follow-up paper.
\begin{figure}[htb]
  \begin{center}
    \epsfig{bbllx=19,bblly=19,bburx=576,bbury=455,file=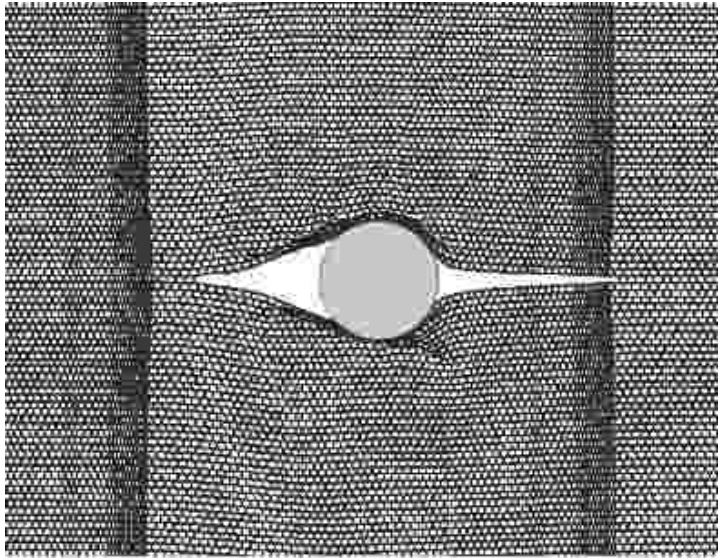, width=9.5cm}
    \caption{Snapshot of the damage for the permeation of a nail.}
  \end{center}
\end{figure} 
\section{Discussion}
A disordered hexagonal beam network model has been introduced to study the damage development in the natural fiber composite wood. The main advantage of our modeling is, that it naturally accounts for the complicated local stress fields formed around failed regions in wood. Furthermore, it captures the gradual activation of the two relevant damage mechanisms and their interaction during the fracture process. We have demonstrated that our discrete element model is capable to provide a deep insight into the damage process occurring perpendicular to grain under gradual loading of wood. Experimental results on the microstructure of damage and on the loss of stiffness are in good agreement with numerical simulations. With our model even dynamics of fracture during the processing of wood can be simulated, as demonstrated on the damage development caused by the permeation of a nail.

An extension of the model to the description of time-dependent failure processes is straight forward, since time-dependent rheological elements can easily be inserted on the microstructural level. The capabilities of our model are not limited to wood, it can be applied to study dynamic damage processes of other cellular or micro-structured materials under varying loading cases including also thermal or chemical degradation processes resulting in complex constitutive governing equations of the single elements. Studies in this direction are in progress.
\section{Acknowledgment} 
The presented work is partly funded by the German Science Foundation (DFG) within the Collaborative Research Center SFB 381 'Characterization of Damage Development in Composite Materials using Non-Destructive Test Methods' under Project C7 'Characterization of Damage Relevance using Scale Spanning Description of Damage Evolution', which is gratefully acknowledged. Henry Gerhard from the IKP, University of Stuttgart is thanked for the CLSM pictures.

\end{document}